\documentclass[fleqn,10pt]{wlscirep}
\usepackage[utf8]{inputenc}
\usepackage[T1]{fontenc}
\usepackage{amsmath,amssymb,amsthm}
\usepackage{xcolor}
\usepackage{tikz}
\usepackage{pgfplots}
\usepackage{svg}
\usetikzlibrary{quantikz}
\usetikzlibrary{patterns,patterns.meta}
\usepackage{rotating}

\definecolor{PictureColor1}{RGB}{15,111,198}
\definecolor{PictureColor2}{RGB}{0,157,217}
\definecolor{PictureColor3}{RGB}{11,208,217}
\definecolor{PictureColor4}{RGB}{16,207,155}
\pgfplotsset{compat=1.17}

\title{Implementation of quantum compression on IBM quantum computers}

\author[1,2]{Matej Pivoluska}
\author[1,2]{Martin Plesch}
\affil[1]{Institute of Physics, Slovak Academy of Sciences, Dúbravská cesta 9, SK-841 04 Bratislava, Slovak Republic}
\affil[2]{Institute of Computer Science, Masaryk University, Šumavská 416, CZ-602 00 Brno, Czech Republic}

\begin{abstract}
Advances in development of quantum computing processors brought ample opportunities to test the performance of various quantum algorithms with practical implementations. 
In this paper we report on implementations of quantum compression algorithm that can efficiently compress unknown quantum information.  
We restricted ourselves to compression of three pure qubits into two qubits, as the complexity of even such a simple implementation is barely within the reach of today's quantum processors.
We implemented the algorithm on IBM quantum processors with two different topological layouts---a fully connected triangle processor and a partially connected line processor.
It turns out that the incomplete connectivity of the line processor affects the performance only minimally. On the other hand, it turns out that the transpilation, i.e. compilation of the circuit into gates physically available to the quantum processor, crucially influences the result. We also have seen that the compression followed by immediate decompression is, even for such a simple case, on the edge or even beyond the capabilities of currently available quantum processors. 
\end{abstract}
\begin{document}

\flushbottom
\maketitle
\thispagestyle{empty}

\section*{Introduction}
Quantum computers, as a theoretical concept, has been suggested in the 1980’s by Richard Feynman. His seminal work \cite{Feynman1982} on simulating quantum physics with a quantum mechanical computer has inspired a new scientific field, collectively known as quantum information and computation \cite{nielsen00}. 
In the last thirty years, the possibility of quantum computing has been studied in depth and revolutionary advances in computation and information science have been made. It has been shown that aside from the ability to simulate quantum physics efficiently, which is invaluable in chemistry \cite{10.3389/fchem.2020.587143,doi:10.1021/acs.chemrev.8b00803}, quantum computers provide a speedup in interesting computational tasks, such as integer factorization \cite{shor}, search in unstructured databases \cite{grover} or random walks \cite{qwalks}. 
Additionally, quantum information scientists have realized that quantum information can be used to implement some communication tasks more efficiently \cite{batmanWinner} and above all else, with unconditional security \cite{QKD}. 

In spite of all these advances, there has always been a large gap between theory and experiments in quantum computation and information. 
While significant theoretical advances---in the form of quantum algorithms and quantum communication protocols---have been achieved already in the 1990’s, implementation of these ideas has, until recently, been lagging. 
Although in quantum computation the most well-known algorithms have obtained a proof-of-principle implementations (see the most recent implementations of Shor’s factorization algorithm \cite{ShorPrinciple} and Grover search \cite{GroverPractical}), researchers were, until recently, commonly unable to test their algorithms even on small scale quantum computers. 
This situation has changed in May 2016, when IBM has made their quantum computers accessible to general public via remote access \cite{IBM}. 
This invigorated the field of quantum computation and since then multiple experiments have been conducted on IBM systems and reported on in literature\cite{PhysRevA.96.022117,PhysRevA.95.032131,DEFFNER2017e00444,Huang2016,Wootton_2017,fedortchenko2016quantum,2017,PhysRevA.95.052339,PhysRevA.94.012314,PhysRevA.94.032329,8622457,Acasiete2020,Zhang2021,bharti2021noisy,oliveira2021quantum}. 

In this paper we join this effort and implement quantum compression algorithm introduced in \cite{compression} and further developed in \cite{PhysRevLett.113.160504,PhysRevLett.116.080501,PhysRevLett.117.090502,PhysRevA.102.032412,Bai_2020,https://doi.org/10.1002/que2.67}. This algorithm is used to compress $n$ identical copies of an arbitrary pure qubit state into roughly $\log(n)$ qubits. Unlike in classical physics, in quantum world a set of identical states represents a valuable resource in comparison to a single copy of such a state. As quantum states cannot be copied \cite{Park1970,Wootters1982,DIEKS1982271} and a single copy provides only a limited information about the state when measured \cite{holevo_2001}, several copies can be utilized for repeated use in follow-up procedures or for a more precise measurement. 

Storing $N$ identical copies of the same state independently is obviously a very inefficient approach. Whereas it is not possible to compress the states in the classical manner (concentrating entropy into a smaller subspace) without measuring the states and disturbing them, laws of quantum mechanics allow to utilize the symmetry of a set of identical states to concentrate all relevant information onto a small, yet not constant subspace. In \cite{compression} we have shown that such a procedure can be done in an efficient way (i.e. using a number of elementary quantum gates scaling at most quadratically with the number of compressed states) and this idea was later utilized with a custom designed quantum experiment \cite{PhysRevLett.113.160504} for the specific case of compressing three identical states of qubits on a subspace of two qubits.

Here we implement the same, simplest non-trivial case, which we call $3\mapsto 2$ compression. Unfortunately, larger number of compressed qubits is beyond the scope of current quantum processors, because the depth of the required circuit becomes impractical.
As we show in the Results section, compression followed by immediate decompression is already for this most simple scenario on the edge of capabilities of IBM processors. 
Scaling up to the next level, i.e.  $4\mapsto 3$ compression, would induce an increase of the number of elementary gates by at least a factor of $5$, which would certainly result into a complete noise in the result. 
Another disadvantage of $4\mapsto 3$ compression is a large redundancy in the target space (three qubits can accommodate information about as many as seven identical states), leaving space for further errors in the decompression. 

\begin{figure}[h]
\centering
\begin{quantikz}
\lstick{\ket{\psi}$_0$}  & \ctrl{1}\gategroup[2,steps=2,style={dashed,
rounded corners, fill=PictureColor4!50, inner xsep=2pt},
background]{{\sc QSWT}} & 
\gate[style = {fill = PictureColor3}]{H} &\qw&
\qw\gategroup[3,steps=4,style={dashed,
rounded corners,fill=PictureColor4!50, inner xsep=2pt},
background]{{\sc Map $q_2$ to $q_0$ and $q_1$}}&\targ[fill=PictureColor2]{}&
\qw&\qw&
\qw&\ctrl{2}\gategroup[3,steps=2,style={dashed,
rounded corners, fill=PictureColor4!50, inner xsep=2pt},
background]{{\sc Disentangle and erase $q_2$}}&\qw&\qw& \qw\rstick[wires=2]{$\ket{\varphi_{1,2}}$}&\\
\lstick{\ket{\psi}$_1$} & \targ[style = {fill = PictureColor2}]{} & \ctrl{-1} & \qw&\qw& \ctrl{-1}&\targ[style = {fill = PictureColor2}]{}&\qw&\qw&\qw&\ctrl{1}&\qw&\qw\\
\lstick{\ket{\psi}$_2$} &\qw&\qw&\qw&\qw&\ctrl{-1}&\ctrl{-1}&\qw&\qw&\gate[style = {fill = PictureColor3}]{U_3(1.91,\pi,\pi)}&\gate[style = {fill = PictureColor3}]{U_3(1.23,0,\pi)}&\qw&\rstick{\ket{0}}\qw
\end{quantikz}
    \caption{Basic Circuit for compression of three qubits into two after a series of optimization comparing to the original results presented in \cite{compression}. QSWT stands for Quantum Schur-Weyl Transform, for details see \cite{PhysRevLett.113.160504}. Before execution on a real quantum processor it needs to be transpiled, i.e. compiled into basis gates. }
    \label{fig:naive_implementation1}
\end{figure}
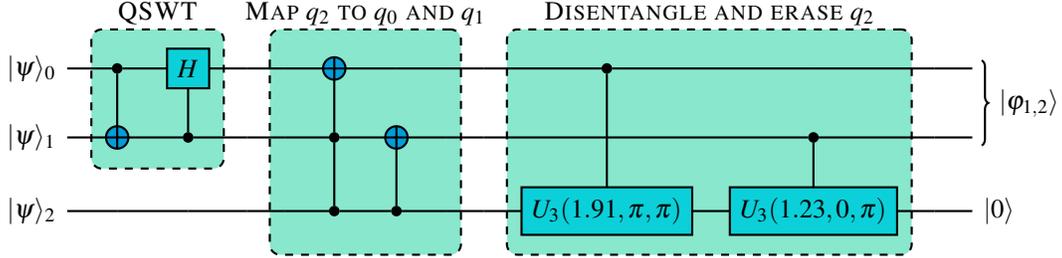

Implemented algorithm can be defined using a gate model of quantum computation and is given in Fig. \ref{fig:naive_implementation1}.
Apart from well known standard gates (CNOT gate, Toffoli gate and controlled $H$ gate) the depicted algorithm uses controlled $U_3$ gates, where 
\begin{center}
 \begin{align}\label{eq:U3}
       U_3(\phi,\theta,\lambda) &= 
    \left[
    \begin{matrix}
        \cos\left(\frac{\theta}{2}\right)& -e^{i\lambda}\sin\left(\frac{\theta}{2}\right)\\
        e^{i\phi}\sin\left(\frac{\theta}{2}\right)&e^{i(\phi+\lambda)}\cos\left(\frac{\theta}{2}\right)
    \end{matrix}
    \right].
 \end{align}
\end{center}
Note that $U_3$ gate is just a specific parametrization of a universal one qubit unitary. Implementing the $3\mapsto2$ compression algorithm is in principle possible simply by inserting the circuit from Fig. \ref{fig:naive_implementation1} into the IBM quantum computing platform called Qiskit\cite{qiskit}, and running it using a simulator or a real processor. 
This, however, rarely leads to an optimal, or even acceptable implementation in terms of fidelity of the compressed  state to the ideal compressed one.
The main reason for this is that controlled $H$ gate, Toffoli gate and the controlled $U_3$ gates cannot be natively executed on the IBM quantum processors and need to be decomposed into the hardware supported \textit{basis gates}. Procedure to perform this decomposition is called \textit{transpilation}. 
The basis gates of IBM  quantum computers are: $R_z(\theta)$ -- a rotation around $z$ axis by an angle $\theta$;  $\sqrt{X}$ -- a square root of Pauli $X$ gate; and CNOT -- a controlled not gate.
The final form of the circuit to be executed is further guided by the connectivity graph of the quantum processor to be used, which contains an information about which pairs of qubits can perform a hardware CNOT operation.
There are only two types of connectivity graphs for a connected configuration of $3$ qubits:(1) a triangle, a fully connected graph in which CNOT can be implemented between all pairs of qubits and (2) a line, in which one CNOT between one pair of qubits is not available. 
These are both relevant for practical quantum computing on IBM quantum platform, as at the time of performing the experiments processors of both kinds were available.

The paper is organized as follows. In the first part, we present the results of simulations and experiments for the compression algorithm only, both on the fully connected quantum processor and on the partially connected processor, where a more sophisticated transpilation is needed. In the second part we present the results of a combined compression and immediate decompression algorithm, both for fully connected and partially connected processors. Here the transpilation takes even a bigger role, as the internal IBM system was not able to fully optimize the circuits, unlike in the previous case, so a custom post-processing lead to better results.

% \begin{figure}[h]
%     \centering
%     \includegraphics[width=0.7\textwidth]{CompressionCircuit.jpg}
%     \caption{Basic Circuit for compression of three qubits into two.}
%     \label{fig:naive_implementation2}
% \end{figure}

\section*{Results}
%As mentioned above, we have implemented the $3\mapsto 2$ compression protocol on IBM quantum processors with different connectivity graphs. We have been using \textit{} --- line and triangle. Since the original algorithm requires a full qubit connectivity, the final transpiled form of the circuit is different on the line processor and on the triangle processor. 
%Because it is unlikely that quantum processors in the near future will contain triangles as their connectivity subgraphs, it is important to investigate how much the line connectivity will impact the performance of the algorithm. 
We conducted two different $3 \mapsto 2$ compression experiments. First, we performed a compression only algorithm, in which we run the compression algorithm and perform the full tomography of the resulting $2$ qubit states to obtain fidelity to the ideal compressed state. Second, we performed compression algorithm followed by decompression algorithm, in which we first compress three input states into two and then proceed to perform the decompression algorithm. This experiment can be seen as a simulation of the whole compression/decompression routine with an assumption of faultless quantum memory. Here we do not need to make a full tomography of the resulting state, as the fidelity is given simply by its $000$ state component. 

As the input state $\ket{\psi}$ significantly affects the fidelities obtained, each of the two experiments was performed on $6$ different input states -- eigenvectors of Pauli $X,Y$ and $Z$ operations denoted $\ket{+},\ket{-},\ket{y_+},\ket{y_-},\ket{0},\ket{1}$.
Further, we implemented each of these experiments in two ways -- one using default calls of transpilation function provided by the IBM programming environment Qiskit\cite{transpile} and the second using a more sophisticated transpilation algorithm, which first splits the compression circuits into subparts and transpiles them separately before one final transpilation as a whole (see Methods section for detailed description).
Transpilation is performed using simulators of quantum processors \textit{ibmq\_5\_yorktown} (triangle connectivity) and {\it ibmq\_bogota} (line connectivity) as backend. Choosing a backend informs the transpilation function about the connectivity and current calibration data which is used in an attempt to find the best decomposition into the elementary quantum gates. 
We show that using more sophisticated transpilation, we can significantly decrease the number of  single- and two-qubit gates needed, which results in decreased depth and increased fidelity in most of the performed experiments (see Figures \ref{fig:TranspiledCircuits}, \ref{fig:CompressionResults}  and \ref{fig:CDresults}).

The first result of this paper is that the implementation on a line connected processor does not require substantially more resources than the fully connected triangle architecture ---
triangle implementation of the compression circuit requires $9$ CNOTs, while the line implementation requires only $10$ CNOTs. Thus the overhead of the incomplete connectivity is limited to about $10\%$ and is compensated by the lower noise of the processor with limited connectivity. 

Finally, we run both experiments with different starting states and using both efficient and default transpilation on real quantum hardware. 
This reveals that the simulators are too optimistic as the decrease in fidelity for all cases is rather significant. 
This effect becomes more pronounced with larger number of gates in the tested circuit, which is apparent from the fact that compression experiment on real hardware produces rather good outcomes even for real hardware (see Figure  \ref{fig:CompressionResults}), while compression/decompression experiment results in very low fidelities of correct decompression (see Figure \ref{fig:CDresults}).

\subsection*{Compression experiment}
\begin{figure}[htp!]
%\centering
\begin{quantikz}[column sep=0.15cm]
\lstick{\ket{\psi}$_0$} & \qw &\gate[style = {fill=PictureColor3}]{U_3}\gategroup[3,steps=19,style={dashed,
rounded corners, fill=PictureColor4!50, inner xsep=2pt},
background]{{\sc Default transpilation for ibmq\_5\_yorktown}} &\ctrl{1}&\gate[style = {fill=PictureColor3}]{U_3}&\ctrl{1}& \gate[style = {fill=PictureColor3}]{U_3}&\targ[fill=PictureColor2]{}&\gate[style = {fill=PictureColor3}]{U_3}&\targ[fill=PictureColor2]{}&\gate[style = {fill=PictureColor3}]{U_3}&\ctrl{2}&\gate[style = {fill=PictureColor3}]{U_3}&\qw&\qw&\ctrl{2}&\gate[style = {fill=PictureColor3}]{U_3}&\ctrl{2}&\gate[style = {fill=PictureColor3}]{U_3}&\qw&\qw&\qw &\meter[style = {fill=PictureColor1}]{}\rstick[wires=2]{$\rotatebox{-90}{ \textsc{Tomography} }$}\\
\lstick{\ket{\psi}$_1$} & \qw &\gate[style = {fill=PictureColor3}]{U_3} &\targ[fill=PictureColor2]{}&\gate[style = {fill=PictureColor3}]{U_3}&\targ[fill=PictureColor2]{}& \gate[style = {fill=PictureColor3}]{U_3}&\qw&\qw&\ctrl{-1}&\gate[style = {fill=PictureColor3}]{U_3}&\qw&\qw&\ctrl{1}&\gate[style = {fill=PictureColor3}]{U_3}&\qw&\qw&\qw&\qw&\ctrl{1}&\gate[style = {fill=PictureColor3}]{U_3}& \qw &\meter[style = {fill=PictureColor1}]{}\\
\lstick{\ket{\psi}$_2$} & \qw &\qw  & \qw &\qw&\qw&\qw&\ctrl{-2}&\gate[style = {fill=PictureColor3}]{U_3}&\qw&\qw&\targ[fill=PictureColor2]{}&\gate[style = {fill=PictureColor3}]{U_3}&\targ[fill=PictureColor2]{}&\gate[style = {fill=PictureColor3}]{U_3}&\targ[fill=PictureColor2]{}&\gate[style = {fill=PictureColor3}]{U_3}&\targ[fill=PictureColor2]{}&\gate[style = {fill=PictureColor3}]{U_3}&\targ[fill=PictureColor2]{}&\qw&\qw& \qw&\trash{\textsc{Trash}}
\end{quantikz}
\vspace{0.5cm}

\begin{quantikz}[column sep=0.15cm]
\lstick{\ket{\psi}$_0$} & \qw& \qw\gategroup[3,steps=17,style={dashed,
rounded corners, fill=PictureColor4!50, inner xsep=2pt},
background]{{\sc Efficient transpilation for ibmq\_5\_yorktown}} & \ctrl{1} &\gate[style = 
{fill = PictureColor3}]{U_3} &\ctrl{1}&\gate[style = 
{fill = PictureColor3}]{U_3}&\targ[style = 
{fill = PictureColor2}]{} &\gate[style = 
{fill = PictureColor3}]{U_3} &\targ[style = 
{fill = PictureColor2}]{}&\gate[style = 
{fill = PictureColor3}]{U_3}&\targ[style = 
{fill = PictureColor2}]{}&\gate[style = 
{fill = PictureColor3}]{U_3}&\qw&\ctrl{2}&\qw&\ctrl{2}&\qw&\qw&\qw&\meter[style = 
{fill = PictureColor1}]{}\rstick[wires=2]{$\rotatebox{-90}{ \textsc{Tomography} }$}\\
\lstick{\ket{\psi}$_1$} &\qw& \gate[style = 
{fill = PictureColor3}]{U_3} &\targ[style = 
{fill = PictureColor2}]{} &\gate[style = 
{fill = PictureColor3}]{U_3}&\targ[style = 
{fill = PictureColor2}]{}&\gate[style = 
{fill = PictureColor3}]{U_3}&\qw&\qw &\ctrl{-1}&\gate[style = 
{fill = PictureColor3}]{U_3}&\qw&\targ[style = 
{fill = PictureColor2}]{}&\gate[style = 
{fill = PictureColor3}]{U_3}&\qw&\qw&\qw&\qw&\ctrl{1}&\qw&\meter[style = 
{fill = PictureColor1}]{}\\
\lstick{\ket{\psi}$_2$} & \qw&\qw & \qw &\qw &\qw&\qw&\ctrl{-2}&\qw&\qw&\qw&\ctrl{-2}&\ctrl{-1}&\gate[style = 
{fill = PictureColor3}]{U_3}&\targ[style = 
{fill = PictureColor2}]{}&\gate[style = 
{fill = PictureColor3}]{U_3}&\targ[style = 
{fill = PictureColor2}]{}&\gate[style = 
{fill = PictureColor3}]{U_3}&\targ[style = 
{fill = PictureColor2}]{}&\qw& \qw&\trash{\textsc{Trash}}
\end{quantikz}
\vspace{0.5cm}

\begin{quantikz}[column sep=0.15cm]
\lstick{\ket{\psi}$_0$} & \qw& 
\gate[style = {fill = PictureColor3}]{U_3}\gategroup[3,steps=21,style={dashed,
rounded corners, fill=PictureColor4!50, inner xsep=2pt},
background]{{\sc Default transpilation for ibmq\_bogota}}  &
\targ[fill = PictureColor2]{} &
\gate[style = {fill = PictureColor3}]{U_3} &
\targ[fill = PictureColor2]{}&
\gate[style = {fill = PictureColor3}]{U_3} &
\qw & \qw&
\ctrl{1} & 
\gate[style = {fill = PictureColor3}]{U_3} &
\qw & \qw & \qw & \qw &
\targ[fill = PictureColor2]{}&
\gate[style = {fill = PictureColor3}]{U_3} &
\qw & \qw & \qw & \qw &
\targ[fill = PictureColor2]{}&
\gate[style = {fill = PictureColor3}]{U_3} &  \qw & \qw &
\meter[style = {fill = PictureColor1}]{}\rstick[wires=3]{$\rotatebox{-90}{ \textsc{Tomography} }$}
\\
\lstick{\ket{\psi}$_1$} & \qw & 
\gate[style = {fill = PictureColor3}]{U_3} &
\ctrl{-1}& 
\gate[style = {fill = PictureColor3}]{U_3} &
\ctrl{-1}&
\gate[style = {fill = PictureColor3}]{U_3} &
\targ[fill = PictureColor2]{}&
\gate[style = {fill = PictureColor3}]{U_3} &
\targ[fill = PictureColor2]{}&
\gate[style = {fill = PictureColor3}]{U_3} &
\targ[fill = PictureColor2]{}&
\gate[style = {fill = PictureColor3}]{U_3} &
\targ[fill = PictureColor2]{}&
\gate[style = {fill = PictureColor3}]{U_3} &
\ctrl{-1}&
\gate[style = {fill = PictureColor3}]{U_3} &
\targ[fill = PictureColor2]{}&
\gate[style = {fill = PictureColor3}]{U_3} &
\targ[fill = PictureColor2]{}&
\gate[style = {fill = PictureColor3}]{U_3} &
\ctrl{-1}& \qw&  \qw& 
 \trash{\textsc{Trash}}&
\\
\lstick{\ket{\psi}$_2$} & \qw &\qw 
&\qw &\qw &\qw&\qw&
\ctrl{-1}&
\gate[style = {fill = PictureColor3}]{U_3} &
\qw & \qw & 
\ctrl{-1}&
\gate[style = {fill = PictureColor3}]{U_3} &
\ctrl{-1}&
\gate[style = {fill = PictureColor3}]{U_3} &
\qw & 
\qw &
\ctrl{-1}&
\gate[style = {fill = PictureColor3}]{U_3} &
\ctrl{-1}&
\qw &\qw &\qw & \qw & \qw &
\meter[style = {fill = PictureColor1}]{} &
\end{quantikz}
\vspace{0.5cm}

\begin{quantikz}[column sep=0.15cm]
\lstick{\ket{\psi}$_0$} & \qw& 
\gate[style = {fill = PictureColor3}]{U_3} \gategroup[3,steps=19,style={dashed,
rounded corners, fill=PictureColor4!50, inner xsep=2pt},
background]{{\sc Efficient transpilation for ibmq\_bogota}}&
\targ[fill = PictureColor2]{}&
\gate[style = {fill = PictureColor3}]{U_3} &
\targ[fill = PictureColor2]{}&
\gate[style = {fill = PictureColor3}]{U_3} &
\qw & \qw &
\ctrl{1}& \qw &
\qw & \qw &
\qw & \qw &
\targ[fill = PictureColor2]{}&
\gate[style = {fill = PictureColor3}]{U_3} &
\qw &
\qw & \qw & 
\ctrl{1} &
\qw & \qw & \qw &
\meter[style = {fill = PictureColor1}]{}\rstick[wires=3]{$\rotatebox{-90}{ \textsc{Tomography} }$} 
\\
\lstick{\ket{\psi}$_1$} & \qw& \qw&
\ctrl{-1}&
\gate[style = {fill = PictureColor3}]{U_3} &
\ctrl{-1}&
\gate[style = {fill = PictureColor3}]{U_3} &
\targ[fill = PictureColor2]{}&
\gate[style = {fill = PictureColor3}]{U_3} &
\targ[fill = PictureColor2]{}&
\gate[style = {fill = PictureColor3}]{U_3} &
\targ[fill = PictureColor2]{}&
\gate[style = {fill = PictureColor3}]{U_3} &
\targ[fill = PictureColor2]{}&
\gate[style = {fill = PictureColor3}]{U_3} &
\ctrl{-1}&
\targ[fill = PictureColor2]{}&
\gate[style = {fill = PictureColor3}]{U_3} &
\targ[fill = PictureColor2]{}&
\gate[style = {fill = PictureColor3}]{U_3} &
\targ[fill = PictureColor2]{}&
\qw&  \qw& 
 \trash{\textsc{Trash}}&
\\
\lstick{\ket{\psi}$_2$} & \qw& \qw&
\qw&
\qw& \qw&
\qw&
\ctrl{-1}&
\gate[style = {fill = PictureColor3}]{U_3} &
\qw & \qw &
\ctrl{-1}&
\gate[style = {fill = PictureColor3}]{U_3} &
\ctrl{-1}&
\gate[style = {fill = PictureColor3}]{U_3} &
\qw &
\ctrl{-1}&
\qw &
\ctrl{-1}&
\qw & \qw & 
\qw & \qw & \qw &
\meter[style = {fill = PictureColor1}]{} 
\end{quantikz}

\caption{Schematic representation of transpiled compression circuits. Here, single qubit rotations labeled $U_3$ (see Eq. \eqref{eq:U3} for definition) are implemented by $1-5$ basis gates, i.e. $R_z$ rotations and $\sqrt{X}$. 
In total, the default transpiled circuit using a triangle architecture contains $9$ CNOTs, $35$ $R_Z$ operations and $28$ $\sqrt{X}$ operations, with depth $46$. 
Efficiently transpiled circuit using triangle architecture consists of $9$ CNOTs, $23$ $R_Z$ and $14$ $\sqrt{X}$ operations and has depth $37$. 
The best circuit produced by default transpiler on a line architecture 
consists of $10$ CNOTs, $30$ $R_Z$ operations and $26$ $\sqrt{X}$ operations with depth $49$.
Efficient transpilation always finds a solution with $10$ CNOT, $24$ $R_Z$ and $18$ $\sqrt{X}$ operations and also has depth $41$. 
These circuits are used in the compression experiment, where full tomography is performed on the qubits containing the compressed state, while the third qubit is discarded. Note that in the circuit for the triangle connectivity qubit $2$ is discarded and for the line connectivity qubit $1$ is discarded.
}
    \label{fig:TranspiledCircuits}
\end{figure}
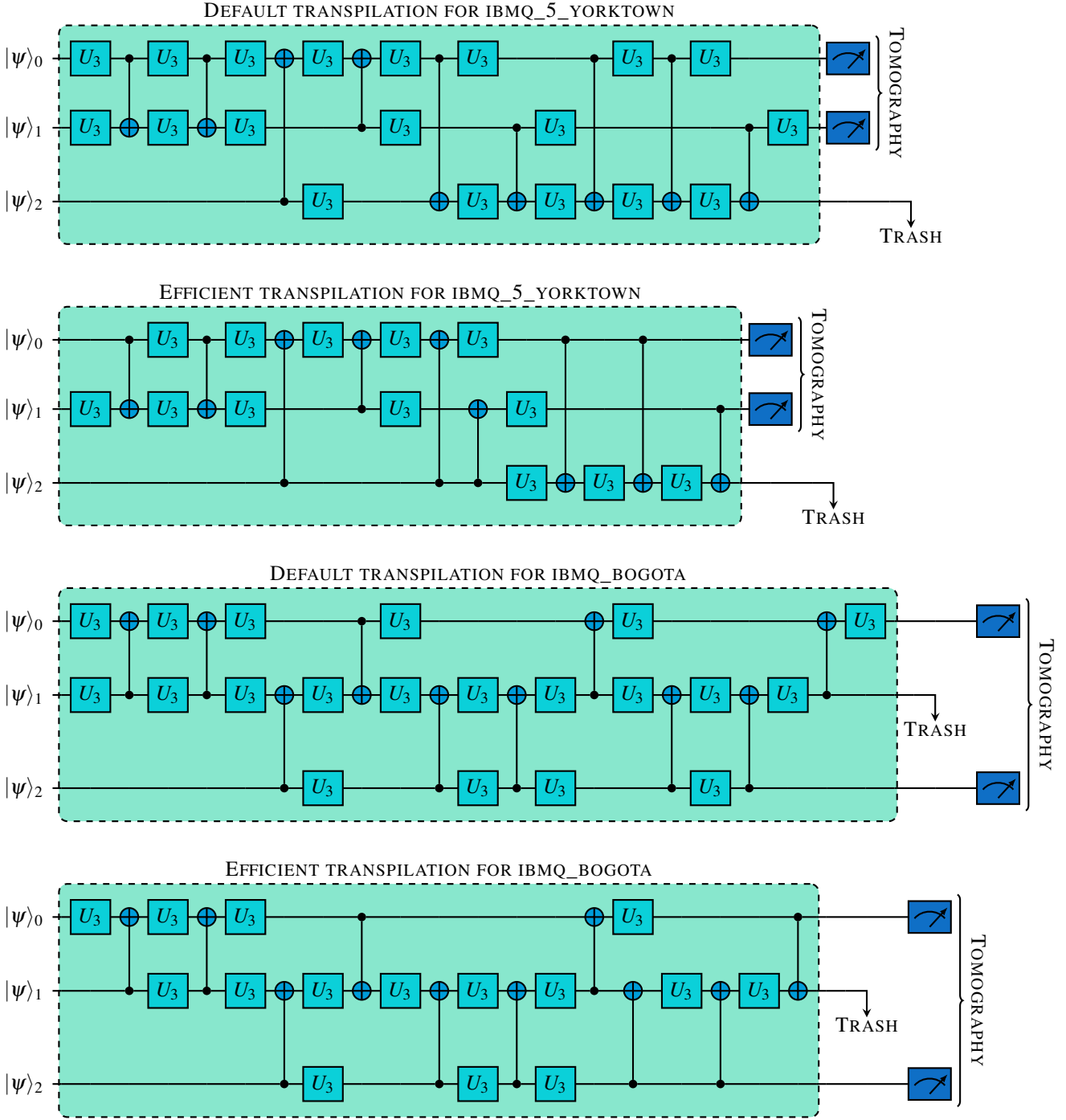

In this subsection we present detailed results for the compression only experiment. First we conducted experiments with triangle connectivity, using \textit{ibmq\_5\_yorktown} quantum processor.
Default transpilation with this backend produces a circuit with $9$ CNOTs, $35$ $R_Z$ operations and $28$ $\sqrt{X}$ operations, with depth $46$. 
The efficient transpiration  results in a transpiled circuit with $9$ CNOTs, $23$ $R_Z$ operations and $14$ $\sqrt{X}$ operations with depth $37$ (see Figure \ref{fig:TranspiledCircuits} for a schematic representation of these circuits).
This difference results in roughly $1-2\%$ increase of fidelity when simulating the efficiently transpiled compression algorithm, except for starting state $\ket{1}$, where the default solution slightly outperforms the efficient one.

In experiments using line architecture we were using \textit{ibmq\_bogota} processor. With this backend the transpilation function produced variable results.
Number of CNOTs varied between $10$ and $25$, while the circuit depth varied between $49$ and $105$. 
The reason for this variance is that transpilation procedure uses a stochastic method to find decomposition in case of a missing CNOT connectivity.
Roughly $25\%$ of runs find the most efficient solution with $10$ CNOTs, $30$ $R_Z$ operations and $26$ $\sqrt{X}$ operations with depth $49$. 
On the other hand, the efficient transpilation (see Methods section for details) resulted in transpiled circuit with $10$ CNOTs, $24$ $R_Z$, and $18$ $\sqrt{X}$ operations and depth $41$  (see Fig. \ref{fig:TranspiledCircuits} for schematic representation of these circuits). 
Surprisingly, this difference results only in negligible increase of fidelity of the simulated compression algorithm when using the efficient solution.
Even more interestingly, in case of input state $\ket{1}$ the default solution again outperforms the efficient one.
\begin{figure}[htp!]
    \centering
   \includegraphics[width=1.125\textwidth]{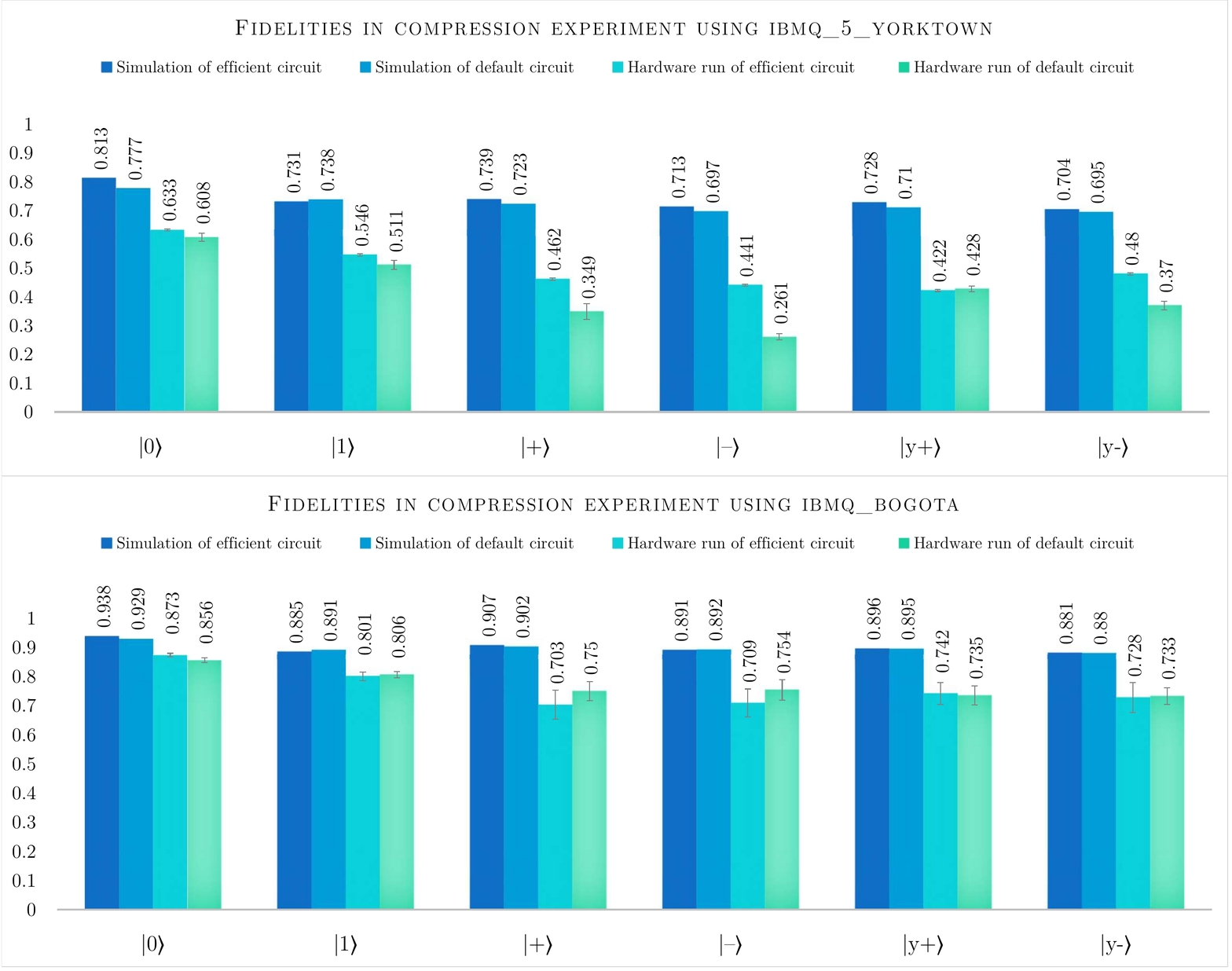}
    \caption{Column chart representing fidelities of the compressed state in the compression experiment using both \textit{ibmq\_5\_yorktown} and \textit{ibmq\_bogota} quantum processors. 
    Simulations were performed with $1$ million shots each, using error parameters provided by IBM for each of the processors, while the hardware run fidelities are averages calculated over $10-20$ runs with $8192$ shots each. We use standard deviation for error bars.}
    \label{fig:CompressionResults}
\end{figure}

In order to confirm the results obtained in simulation we also ran the same circuits on real hardware. 
For \textit{ibmq\_5\_yorktown} the obtained fidelities are  significantly lower than the simulation suggests, with an average drop of $20-30\%$ and in case of default transpiled circuit, ranging even to $40\%$  for the  $\ket{-}$ input state (see Figure \ref{fig:CompressionResults}). 
The length of the circuit is the likely reason for this decrease, as the coherence times for the \textit{ibmq\_5\_yorktown} processor were rather short compared to the newer generation of processors.  
This intuition is confirmed by inspecting the experiment results for \textit{ibmq\_bogota}.
In this set of experiments with a newer generation of the processor with longer coherence times the fidelity decrease compared to the simulation was only $7-20\%$ (see Figure \ref{fig:CompressionResults}). 
Interestingly, also in this case we can see that the best default transpiled circuit outperformed the efficient one for some input states.

\subsection*{Compression and decompression experiment}
\begin{figure}[htp]
    \centering
    \begin{quantikz}[slice all,remove end slices=1, slice titles = {} ,slice style=PictureColor1,slice label style
={inner sep=1pt,anchor=south west,rotate=40}]
    \lstick{\ket{0}$_0$}  & 
    \gate[style = {fill = PictureColor3}]{U_{Prep}}\gategroup[3,steps=5,style={dashed,
rounded corners, fill=PictureColor4!50, inner xsep=2pt},
background]{{\sc Compression and decompression experiment}}& 
    \gate[wires = 3, style = {fill = PictureColor2}]{U_{Comp}} &
    \qw &
    \gate[wires = 3, style = {fill = PictureColor2}]{U_{Comp}^\dagger} &
    \gate[style = {fill = PictureColor3}]{U_{Prep}^\dagger}& 
    \meter[style = {fill = PictureColor1}]{}
    \\
    \lstick{\ket{0}$_1$}  & 
    \gate[style = {fill = PictureColor3}]{U_{Prep}} &
    &
    \qw& 
    &
    \gate[style = {fill = PictureColor3}]{U_{Prep}^\dagger}& 
    \meter[style = {fill = PictureColor1}]{}
    \\
    \lstick{\ket{0}$_2$}  & 
    \gate[style = {fill = PictureColor3}]{U_{Prep}} &
    &
    \gate[style = {fill = PictureColor3}]{\ket{0}} & 
    &
    \gate[style = {fill = PictureColor3}]{U_{Prep}^\dagger}& 
    \meter[style = {fill = PictureColor1}]{}
    \end{quantikz}
    \caption{Schematics of the compression and decompression experiment. Three copies of input state $\ket{0}$ are first prepared into one of the desired starting states $\ket{\psi}$ from the set $\{\ket{0},\ket{1},\ket{+},\ket{-},\ket{y_+},\ket{y_-}\}$ using the preparation unitary $U_{Prep}$. 
    Subsequently, three copies of $\ket{\psi}$ are compressed using the compression algorithm $U_{Comp}$. 
    Then, the last qubit is restored to state $\ket{0}$. This marks the end of the compression part, after which the first two qubits plus the $\ket{0}$ state are first decompressed using the complex conjugation of $U_{Comp}$, and rotated using the complex conjugation of the preparation unitary. 
    The expected result is $\ket{000}$ state and the probability of obtaining this result is the fidelity of the experimental compression and decompression experiment. 
    Dashed vertical lines represent barriers that divide the circuits into parts, which the transpiler processes separately and independently.}
    \label{fig:CDexperiment}
\end{figure}
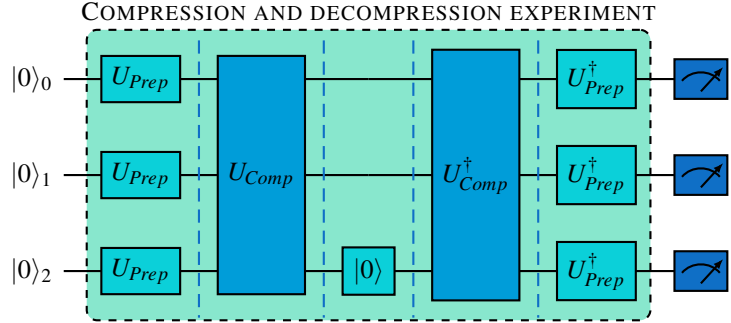

\begin{figure}[htp!]
    \centering
    \includegraphics[width=\textwidth]{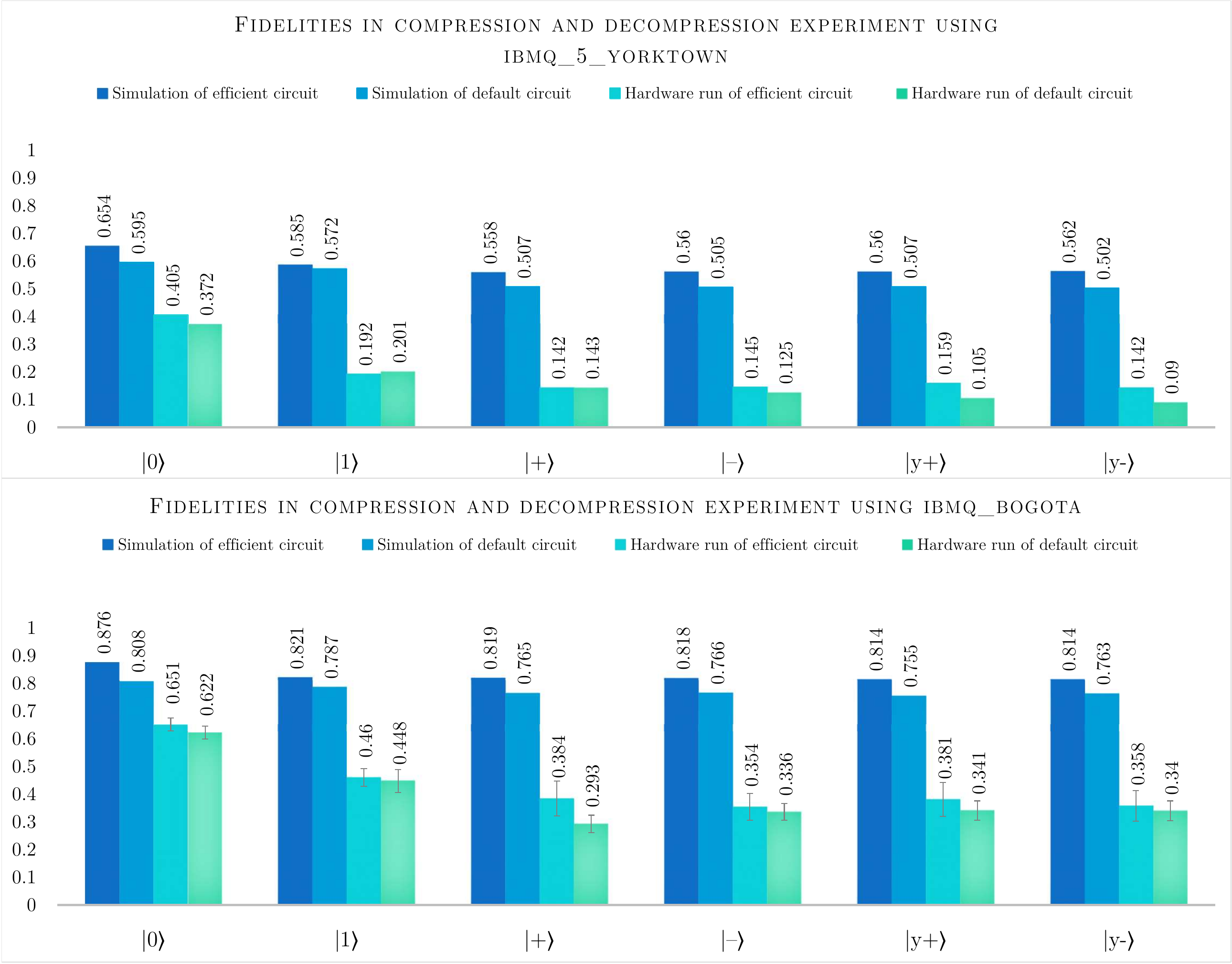}
    \caption{Column chart representing results of the compression and decompression experiment. We plot the fidelity of the decompressed states to  $\ket{\psi}^{\otimes 3}$, where $\ket{\psi}$ is the input state. 
    Simulations were performed with $1$ million shots each, using error parameters provided by IBM for each of the hardware backends (\textit{ibmq\_5\_yorktown} and \textit{ibmq\_bogota}). Due to the expected low fidelity on \textit{ibmq\_5\_yorktown}, the experiments were run only once with $8192$ shots. On the other hand, in case of hardware run on \textit{ibmq\_bogota} we present averages calculated over $10$ runs with $8192$ shots each. We use standard deviation for error bars.}
    \label{fig:CDresults}
\end{figure}

In the second experiment we let the default transpiler produce the circuits for both compression and decompression (see Figure \ref{fig:CDexperiment}). 
In case of \textit{ibmq\_5\_yorktown} the default transpiler could not find a decompression circuit with $9$ CNOTs and the complete compression/decompression circuit therefore had $21$ CNOTs, $62$ $R_Z$ operations and $46$ $\sqrt{X}$ operations with total depth of $90$ including the operation, which resets the third qubit. 
This compares to efficient circuit for triangle connectivity, which uses complex conjugate of the efficient compression algorithm for decompression with $18$ CNOTs, $46$ $R_Z$ and $28$ $\sqrt{X}$ with total depth of $77$ including the reset operation. 
This more significant difference results in larger advantage of simulating the efficient solution compared to the compression only experiment, with efficient circuit reaching roughly $5-6\%$ better fidelities. 
Again, the outlier is the prepared state $\ket{1}$, where the advantage of the efficient solution is only roughly $1\%$ (see Figure \ref{fig:CDresults}).

Using \textit{ibmq\_bogota} with line connectivity, we again observe that decompression algorithm poses a problem for the default transpiler. 
The solutions vary considerably with circuits using between $26$ and $49$ CNOTs with depths between $107$ and $201$.
In this experiment only roughly $3\%$ of transpiler runs resulted in the best solution with $26$ CNOTs, $69$ $R_Z$ and $60$ $\sqrt{X}$ operations with total depth of $99$ including the three reset operations.
For comparison, the efficient solution always results in $20$ CNOTs, $48$ $R_Z$ and $36$ $\sqrt{X}$ operations with total depth of $87$.
This decrease in complexity results in advantage for the simulated efficient solution with average increased fidelity between $3-7\%$, depending on the input state.

Moving on to experiments with hardware backend, we see that \textit{ibmq\_5\_yorktown} suffers from a substantial performance drop. 
In particular, the results in the default transpilation case are consistent with random outcomes, suggesting the experimental state fully decohered before the calculation could finish.
Similarly, in the case of experiments using line connected \textit{ibmq\_bogota} backend we can observe a significant drop in fidelities. 
The difference between default and efficient circuits is more substantial than in the compression only experiment, which is caused by a more substantial difference between the two circuits. 
Here, clearly the efficient solution outperformed the default one, however, both suffered $20-45\%$ drop in obtained fidelities comparing to simulations.
This suggests that the length of the circuits currently exceeds the possibilities of even the newest generation of IBM quantum computers.

\section*{Discussion}

{Compression of unknown quantum information in its simplest scenario, compressing of three identical states into two, is a nice toy example for testing of abilities of emerging quantum computers. }
In this work we present the implementation of the quantum compression algorithm on two different IBM processors. In both cases we simulate the procedure using classical computers and run real quantum computations.

The first result is the comparison of two different types of quantum processor connectivity---full triangle and a line connectivity.
Our implementations reveal that triangle connectivity does not result in a significant advantage for the $3\mapsto 2$ compression, as only one additional CNOT is needed to compensate the missing connection. 
In other words, the higher quality of the newer generation of processors fully compensated the lower level of connectivity. 
As a result, we have seen that the most recent generation of IBM quantum processors can attain fidelity of compression of $70\%-87\%$, depending on the state to be compressed. 
On one hand this is a rather impressive technical feat, because the implemented circuits are non-trivial, on the other,  it is still likely below the levels needed for practical use of the compression algorithm.

There are also several results that have a general validity for basically any computation performed on quantum computers.
First, we have shown that the current qiskit transpiler needs to be used wisely, with some sophistication. 
This is demonstrated by the fact that the default setting of the transpiler finds the most efficient solution in case of line connectivity with only a very small probability. 
Even worse, the default transpiler does not find the best solution for decompression circuit at all. 
This suggests that in order to find optimal transpiled circuits for any algorithm described by a unitary $U$, it generally might be a good strategy to transpile both $U$ and $U^\dagger$ and choose the more efficient one. 
{It also turns out that it is advantageous to transpile more complicated circuits first in smaller blocks to get rid of unsupported gates and connections and then optimize the whole circuit in order to minimize the total number of gates. } 

As a very important point it turns out that the simulators implemented for IBM quantum computers are far too optimistic. Most probably only a part of the decoherence sources is sufficiently modeled, which leads to a far lower simulated noise if compared to reality. 
This in particular limits its usability for testing the performance of the available processors on complicated tasks.

\section*{Methods}
In this section we briefly describe the tools and the algorithm used to obtain efficient circuits. 
The main tool {that crucially influenced the quality of the output results} was \texttt{transpile} function from qiskit. {It translates all gates that are not directly supported by the computer into gates from its library and also bridges CNOT gates acting between not-connected qubits into a series of gates on connected qubits. It should, to some level, also optimize the circuit for the least number of gates and get use of the higher quality qubits and connections. }

In its basic form, \texttt{transpile} function takes as inputs \texttt{circuit}, \texttt{backend} and \texttt{optimization\_level}. 
Input \texttt{circuit} contains the information about the circuit to be transpiled and \texttt{backend} contains information about the quantum processor to be used -- connectivity of given quantum processor (i.e. line or triangle in our case), as well as error parameters of individual qubits.
The last input defines what kind of optimization is performed on the circuit. There are four basic levels qiskit offers, described in the Qiskit tutorial \cite{} as:

\begin{itemize}
\item[]
\texttt{optimization\_level=0:} just maps the circuit to the backend, with no explicit optimization (except whatever optimizations the mapper does).
\item[]
\texttt{optimization\_level=1:} maps the circuit, but also does light-weight optimizations by collapsing adjacent gates.
\item[]
\texttt{optimization\_level=2:} medium-weight optimization, including a noise-adaptive layout and a gate-cancellation procedure based on gate commutation relationships.
\item[]
\texttt{optimization\_level=3:} heavy-weight optimization, which in addition to previous steps, does resynthesis of two-qubit blocks of gates in the circuit.
\end{itemize}

{For all settings, the approach of the transpiler is stochastic. Thus it does not necessarily ends up with the same solution every time it is called. Moreover, for more complicated circuits and higher optimization levels it might not find a solution at all, most probably due to reaching a threshold in the number of iterations or computer load.}

The default circuits for both experiments with \textit{ibmq\_5\_yorktown} were obtained by transpiling the default circuit presented in Fig.\ref{fig:naive_implementation1} three times in a row, each time with decreasing the value of the optimization level, starting from value $3$.
The default circuits for \textit{ibmq\_bogota} were obtained in the same way. The transpilation was run $100$ times for both experiments and the most efficient circuits were used.

The efficient circuits for compression experiment were obtained by first splitting circuit presented in Fig. \ref{fig:naive_implementation1} into three parts, in order to transpile both controlled $U_3$ operations and the ``Disentangle and erase'' part separately. First this was done using \textit{ibmq\_5\_yorktown}, with \texttt{optimization\_level = 3} for $U_3(1.23,0,\pi)$ and \texttt{optimization\_level = 2} for $U_3(1.91,\pi,\pi)$, to produce a circuit with $1$ CNOT and $2$ CNOTs respectively. Then the rest of the circuit from Fig.\ref{fig:naive_implementation1} was transpiled with  \texttt{optimization\_level = 1}. Finally all three parts were joined together and again transpiled with  \texttt{optimization\_level = 3}, followed by \texttt{optimization\_level = 1} to produce the final result.

In order to produce an efficient circuit for \textit{ibmq\_bogota}, the efficient circuit for \textit{ibmq\_yorktown} was transpiled three times with a new backend, with $optimization\_level$ starting at value $3$, followed by value $2$ and finally value $1$. Again, the result of this transpilation was stochastic, but in roughly $10\%$ of the trials the final circuit with $10$ CNOTS was produced.

Efficient circuits for the compression and decompression experiment were obtained by using previously obtained efficient circuits for compression experiment and their complex conjugation for decompression part.

\section*{Acknowledgements}

We acknowledge the support of VEGA project 2/0136/19 and GAMU project MUNI/G/1596/2019.
Further, we acknowledge the use of IBM Quantum services for this work. The views expressed are those of the authors, and do not reflect the official policy or position of IBM or the IBM Quantum team. We acknowledge the access to advanced services provided by the IBM Quantum Researchers Program. 

\section*{Author contributions statement}
{Both authors designed the experiment based on the algorithm developed by M.Pl., M.Pi. implemented the programs and both authors analyzed the data. M.Pi. prepared the first draft of the manuscript and both authors edited and finalized it.}

\section*{Additional information}

Authors declare no compering interests.

\end{document}